\documentclass[english,12pt]{article}
\usepackage[cp1251]{inputenc}
\usepackage{babel}
\usepackage{amsfonts, amsmath}

\newcommand{\be}{\begin{equation}} \newcommand{\ee}{\end{equation}}

\begin{document}
\title{Pure States, Mixed States and Hawking Problem
in Generalized Quantum Mechanics} \thispagestyle{empty}

\author{A.E.Shalyt-Margolin\hspace{1.5mm}\thanks
{Fax: (+375) 172 326075; e-mail:alexm@hep.by; a.shalyt@mail.ru;}}
\date{}
\maketitle
 \vspace{-25pt}
{\footnotesize\noindent National Center of Particles and High
Energy Physics, Bogdanovich Str. 153, Minsk 220040, Belarus\\
{\ttfamily{\footnotesize
}}

\rm\normalsize \vspace{0.5cm}
\begin{abstract}
This paper is the continuation of a study into the information
paradox problem started by the author in his earlier works. As
previously, the key instrument is a deformed density matrix in
Quantum Mechanics of the Early Universe. It is assumed that the
latter represents Quantum Mechanics with Fundamental Length. It is
demonstrated that the obtained results agree well with the
canonical viewpoint that in the processes involving black holes
pure states go to the mixed ones in the assumption that all
measurements are performed by the observer in a well-known Quantum
Mechanics. Also it is shown that high entropy for Planck’s
remnants of black holes appearing in the assumption of the
Generalized Uncertainty Relations may be explained within the
scope of the density matrix entropy introduced by the author
previously. It is noted that the suggested paradigm is consistent
with the Holographic Principle. Because of this, a conjecture is
made about the possibility for obtaining the Generalized
Uncertainty Relations from the covariant entropy bound at high
energies in the same way as R.Bousso has derived Heisenberg’s
uncertainty principle for the flat space.
\end{abstract}
\newpage
\section{Introduction}
The present paper is a continuation of analysis of the Hawking
information paradox started by the author in
\cite{sh1}-\cite{sh3}.In these works the above problem has been
considered using a new object introduced by the author in
\cite{sh4} – a deformed density matrix in Quantum Mechanics of the
Early Universe (at Planck scale). The principal idea is as
follows: there is difference between Quantum Mechanics on the
conventional scales and that at Planck’s. In the first case we
have a well-known Quantum Mechanics (QM), and in the second case
we consider Quantum Mechanics with Fundamental Length (QMFL). And
hence in the second case a density matrix differs from the
standard quantum-mechanical one, being a deformation of the latter
through the dimensionless quantity $\alpha=l_{min}^{2}/x^{2}$,
where $l_{min}$ is a minimal length, $x$ is the measurement scale
and the variation interval $\alpha$ is finite $0<\alpha\leq1/4$
\cite{sh5}-\cite{sh8},\cite{sh4}. The deformation is understood as
an extension of a particular theory by inclusion of one or several
additional parameters in such a way that the initial theory
appears in the limiting transition.
\\ In the above-mentioned works an asymmetrical
entropy density matrix depending on two values of the parameter
$\alpha$: $\alpha_{1}$ and $\alpha_{2}$ for the observer and for
the observed has been introduced naturally. Introduction of such a
matrix is important in two ways: first, it enables demonstration
of a possible drastic increase in the entropy density close to the
singularity, providing a new approach to solution of the
information paradox problem \cite{r16}-\cite{r18}; and second, it
indicates that entropy is a relative notion in a sense that it
depends both on the energies  at which an observed object is
measured and on  the fact to what scale, associated with $\alpha$,
belongs the observer. \\ And in \cite{sh3} the author holds the
viewpoint, similar to that of R.Penrose \cite{pen1}, concerning an
increase in entropy close to the singularity: as the entropy
density is the same for the origin (Big Bang) and final
(singularity of Black Hole or Big Crunch) singularities, there is
no information loss at all, i.e. Hawking’s problem is solvable
positively.
\\ In the present paper it is demonstrated that with the developed formalism
the information paradox problem may be solved in principle
 positively using S.Hawking \cite{r16}-\cite{r18} approach as well,
when information is lost during the processes associated with the
horizon of events of a black hole. And with the use of the
proposed approaches it is shown that in processes involving black
holes the initial pure states are always replaced by the mixed
ones near the horizon of events of the black hole. As this takes
place, the results \cite{r16} for thermal radiation of black holes
are not used. (Note that these results have been open to question
recently(for example \cite{Helf})). In this case it is assumed
that all measurements are made by the observer within QM, i.e. by
the observer whose deformation parameter takes the only value
$\alpha_{1}=0$ in conformity with the canonical viewpoint
\cite{r16}, as the primary result of \cite{r16} has been
formulated in a semiclassical approximation corresponding to this
value $\alpha_{1}$.
\\In the second section some tentative results are given
together with important refinements needed later on. The third
section presents the primary result that within the generalized
Quantum Mechanics Hawking’s problem  in a canonical formulation
\cite{r16}-\cite{r18} may be solved positively. In the last
section the author attempts at elucidation of the problem put
forward by J.Bekenstein \cite{bek1} in regard to entropy for
Planck’s remnants of black holes within the proposed paradigm that
is adequately consistent with the holographic principle. Just in
this paper \cite{bek1} the problem is stated as follows: provided
in the process of the black hole evaporation the masses of the
formed remnants are on the order of Planck’s mass $M_{p}$, these
remnants are hardly characterized by high entropy. Whereas in the
developed formalism this is possible due to a high entropy density
at Planck scales, that is measured with the entropy density matrix
$S^{\alpha_{1}}_{\alpha_{2}}$ for $\alpha_{2}$ close to 1/4. As
shown in \cite{sh4}, the relevant matrix element acts as a density
of entropy per unit minimum area depending on the scales for the
observer and observable. In this way for all the results obtained
the principal object under study is the entropy density matrix
introduced previously in \cite{sh1},\cite{sh3} and considered in
Section 3 of the present work.
 In conclusion, as regards the holographic principle, a
conjecture is made about the possibility for derivation of the
Generalized Uncertainty Relations from the covariant entropy bound
at high energies in the same way as R.Bousso \cite{bou2}  has
obtained the Heisenberg uncertainty principle for the flat space.

\section {Quantum Mechanics with Fundamental Length and Density Matrix}

It should be recalled that in \cite{sh4}-\cite{sh8} a new object
has been introduced – a deformation of the density matrix
$\rho(\alpha)$  in QMFL. Here parameter
$\alpha=l_{min}^{2}/x^{2}$, where $l_{min}$ , is a minimal length,
$x$ is the scale of measurement.
\\ The primary properties of $\rho(\alpha)$ are as follows \cite{sh4}.
\\
\noindent {\bf Definition}
\\
\noindent Any system in QMFL is described by a density pro-matrix
of the form $${\bf
\rho(\alpha)=\sum_{i}\omega_{i}(\alpha)|i><i|},$$ where
\begin{enumerate}
\item $0<\alpha\leq1/4$.
\item The vectors $|i>$ form a full orthonormal system;
\item $\omega_{i}(\alpha)\geq 0$ and for all $i$  the
finite limit $\lim\limits_{\alpha\rightarrow
0}\omega_{i}(\alpha)=\omega_{i}$ exists;
\item
$Sp[\rho(\alpha)]=\sum_{i}\omega_{i}(\alpha)<1$,
$\sum_{i}\omega_{i}=1$.
\item For every operator $B$ and any $\alpha$ there is a
mean operator $B$ depending on  $\alpha$:\\
$$<B>_{\alpha}=\sum_{i}\omega_{i}(\alpha)<i|B|i>.$$
\item the following condition must be fulfilled:
\begin{equation}\label{U1}
Sp[\rho(\alpha)]-Sp^{2}[\rho(\alpha)]\approx\alpha.
\end{equation}
\item This suggests limitation for the parameter $\alpha$:
\begin{equation}\label{U2}
Sp[\rho(\alpha)]\approx\frac{1}{2}+\sqrt{\frac{1}{4}-\alpha}.
\end{equation}
\end{enumerate}
In the limit $\lim\limits_{\alpha\rightarrow 0}\rho(\alpha)=\rho$
a well-known QM density matrix appears.
\\ It should be noted that in \cite{sh4}-\cite{sh8} the problems
have been considered with reference to the Generalized Uncertainty
Relations (GUR) \cite{r1}-\cite{r6}, since based on the latter one
arrives to the presence of the fundamental length in a quantum
theory \cite{r6},\cite{Gar}. However, there is no necessity to use
GUR directly for studies of QMFL as all modern approaches to
quantum gravity in some or other way lead to the availability of a
minimal length \cite{Gar}.
\\ Besides, at least heuristically, a theory of Big Bang should
lead to the notion of a minimal length. The reasoning is as
follows:  since for its realization the Big Bang should possess
some maximum energy $E_{max}$ (which may be theoretically
calculated for any specific inflation model), this energy  places
an upper limit for the particle momentum and hence some lower
limit for the length $l_{min}$ even on the basis of the
conventional Heisenberg's Uncertainty Relations \cite{Heis}.
Introduction of GUR \cite{r1}-\cite{r6} just renders this limit
more precise, fixing it at Planck’s level $l_{min}\sim l_{p}$.

\section {Entropy Density Matrix, Pure and Mixed States}

By the proposed approach, in QMFL there are no pure states in a
sense \cite{sh4} that the states with a unit probability can
appear only in the limit $\alpha\rightarrow 0$, when all
$\omega_{i}(\alpha)$ except one are equal to zero or when they
tend to zero in this limit. In our treatment pure states are the
states which can be represented in the form $|\psi><\psi|$, where
$<\psi|\psi>=1$.  That is pure states may appear in the limiting
transition from QMFL to QM only, when the density matrix is
$\lim\limits_{\alpha\rightarrow 0}\rho(\alpha)=\rho$.
\\ At the same time, for every solution of equation (\ref{U1})
one can describe the states going in this limit to the pure ones
in QM. These states may be referred to as prototypes of pure
states. Specifically, such states have been used in \cite{sh4} for
the derivation of a well-known Bekenstein-Hawking formula of the
black hole entropy in a semiclassical approximation. In the
process for one of the solutions of equation (\ref{U1}) the author
has used {\bf exponential ansatz}:
\begin{equation}\label{U3}
\rho^{*}(\alpha)=\sum_{i}\alpha_{i} exp(-\alpha)|i><i|,
\end{equation}
where all $\alpha_{i}>0$ are independent of $\alpha$  and their
sum is equal to 1. In this case the prototype of a pure state
$\rho=|\psi><\psi|$ in QM is a density pro-matrix
$\rho(\alpha)=exp(-\alpha)|\psi><\psi|$ in QMFL. Similarly, the
prototype of any mixed state in QM with the use of an exponential
ansatz is a density pro-matrix, the general form of which is given
by (\ref{U3}).
\\ In \cite{sh1},\cite{sh3}the following entropy density matrix
has been introduced:
\begin{equation}\label{U4}
S^{\alpha_{1}}_{\alpha_{2}}=-Sp[\rho(\alpha_{1})\ln(\rho(\alpha_{2}))]=
-<\ln(\rho(\alpha_{2}))>_{\alpha_{1}},
\end{equation}
where $0< \alpha_{1},\alpha_{2}\leq 1/4.$.
\\ As indicated in \cite{sh1},\cite{sh3},
$S^{\alpha_{1}}_{\alpha_{2}}$ has a clear physical meaning: the
entropy density is computed  on the scale associated with the
deformation parameter $\alpha_{2}$ by the observer who is at a
scale corresponding to the deformation parameter $\alpha_{1}$.
Also note that the matrix $S^{\alpha_{1}}_{\alpha_{2}}$ is always
meaningful irrespective of the fact, whether $\rho(\alpha_{1})$
and $\rho(\alpha_{2})$ form a prototype of pure or mixed state in
QM.
\\Different approaches are taken to the information loss problem
and unitarity violation in the black holes. Specifically,
R.Penroze in \cite{pen1} has demonstrated that information in the
black hole may be lost and unitarity may be violated because of
the singularity. In paper \cite{sh3} it has been stated that in
this case the information loss problem is solved positively. In
the process the entropy density matrix
$S^{\alpha_{1}}_{\alpha_{2}}$ has been primarily used for the
situations when $\rho(\alpha_{2}=1/4$ is the case(being associated
with the singularity). Also in this section we point to a
possibility for solving the problem in principle positively with
the use of the canonical Hawking approach \cite{r16}-\cite{r18},
i.e. for {\bf information losses near the horizon of events} of
the black hole. And independently of the results presented in
\cite{r16}-\cite{r18}, from the developed formalism it follows
that the state measured near the horizon of events is always
mixed. In \cite{r16}-\cite{r18} this is established in view of
thermal radiation exhibited by the black hole. However, in the
last few years the fact of the existence of such a radiation is
open to dispute (e.g., see \cite{Helf}). Therefore, we exclude
this fact from our consideration. Using the developed formalism,
one is enabled to arrive to this result with the above-mentioned
entropy density matrix. Actually, in recent works (e.g.,
\cite{Helf},\cite{Hooft}), it has been shown that near the horizon
of events the quantum-gravitational effects are considerable.
Proceeding from the entropy density matrix, this means that for
the matrix element $S^{\alpha_{1}}_{\alpha_{2}}$ we always have
$\alpha_{2}>0$ as the quantum-gravitational effects are affecting
small scales only.
\\ Now we analyze a random matrix element $S^{\alpha_{1}}_{\alpha_{2}}$.
Obviously, this element may be zero only for the case when
$\rho(\alpha_{2})$ is a pure state measured in QM, i.e.
$\alpha_{2}\approx 0$. A partial case of this situation has been
considered in \cite{r16}, where $\alpha_{1}=\alpha_{2}\approx 0$
and hence $\rho(\alpha_{1})=\rho(\alpha_{2})=\rho_{in}$ with zero
entropy
\begin{equation}\label{U5}
S^{0}_{0}=-Sp[\rho_{in}\ln(\rho_{in})]=0.
\end{equation}
Actually, this is the initial state entropy measured in the
original statement of the information paradox problem
\cite{r16}-\cite{r18}:
\begin{equation}\label{U6}
S^{in}=S^{0}_{0}=-Sp[\rho_{in}\ln(\rho_{in})]=
-Sp[\rho_{pure}\ln(\rho_{pure})]=0.
\end{equation}
The question is, what can be measured by an observer at the exit
for $\rho_{out}$, when all measurements are performed in QM, i.e.
when we have at hand only one deformation parameter
$\alpha_{1}\approx 0$? Simply this means that in QMFL due to the
quantum-gravitational effects at a horizon of the above events in
the black hole \cite{Helf},\cite{Hooft} leads to:
\begin{equation}\label{U7}
S^{out}=S^{\alpha_{2}}_{\alpha_{2}}=-Sp[\rho_{out}\ln(\rho_{out})]=
-Sp[\rho_{\alpha_{2}}\ln(\rho_{\alpha_{2}})]\neq 0
\end{equation}
for some $\alpha_{2}>0$,unapproachable in QM, is conforming to a
particular mixed state in QM with the same entropy
\begin{equation}\label{U8}
S^{out}=-Sp[\rho_{out}\ln(\rho_{out})]=
-Sp[\rho_{mix}\ln(\rho_{mix})]\neq 0.
\end{equation}
It should be emphasized that a mixed state in (\ref{U8}) will be
not uniquely defined. Of course, in this statement there is
information loss, since
\begin{equation}\label{U9}
\Delta S=S^{out}-S^{in}>0.
\end{equation}
However, as shown in \cite{sh1},\cite{sh3}, it will be more
correct to consider the state close to the origin singularity (or
in the early Universe where the quantum-gravitational effects
should be also included) as an initial state for which $S^{in}$ is
calculated, naturally with certain $\alpha>0$ and with entropy
\begin{equation}\label{U10}
S^{in}=S^{\alpha}_{\alpha}=-Sp[\rho_{in}\ln(\rho_{in})]=
-Sp[\rho_{\alpha}\ln(\rho_{\alpha})]>0.
\end{equation}
Again for the observer making measurements in QM and having no
access to any $\alpha>0$ (i.e. having access to $\alpha\approx 0$
only) this is associated with a certain mixed state
\begin{equation}\label{U11}
S^{in}=-Sp[\rho_{in}\ln(\rho_{in})]=
-Sp[\rho_{mix}\ln(\rho_{mix})]>0
\end{equation}
that is also ambiguously determined. In this way the
superscattering operator determined in \cite{r16}-\cite{r18}
\\
$$\$:\rho^{in}\rightarrow\rho^{out}\enskip or
\enskip\$:\rho_{pure}\rightarrow\rho_{mix}$$
\\
in case under consideration will be of the form
\\
$$\$:\rho_{mix}\rightarrow\rho_{mix}.$$
\\
Since in this case $S^{in}>0$ and $S^{out}>0$, may be no
information loss, then
\begin{equation}\label{U12}
\Delta S=S^{out}-S^{in}=0.
\end{equation}
The following points of particular importance should be taken into
consideration:
\\1) A study of the information paradox problem in the generalized
Quantum Mechanics (QMFL) provides extended possibilities for
interpretation of the notion of entropy. Indeed, in the classical
problem statement \cite{r16}-\cite{r18}
$S^{in}=-Sp[\rho_{in}\ln(\rho_{in})]$ is compared with
$S^{out}=-Sp[\rho_{out}\ln(\rho_{out})]$,i.e. within the scope of
the above-mentioned entropy density $S^{\alpha_{1}}_{\alpha_{2}}$
, introduced in \cite{sh1}, \cite{sh3}, two different diagonal
elements $S^{0}_{0}$ and $S^{\alpha}_{\alpha}$ are compared.
However, in the paradigm under consideration one is free to
compare $S^{0}_{0}$ to $S^{0}_{\alpha}$ or
$S^{in}_{in}=-Sp[\rho_{in}\ln(\rho_{in})]$ to
$S^{in}_{out}=-Sp[\rho_{in}\ln(\rho_{out})].$
\\
\\2)Proceeding from this observation and from the results
of \cite{sh1},\cite{sh3}, we come to the conclusion that the
notion of entropy is {\bf relative} within the generalized Quantum
Mechanics (QMFL) in a sense that it is dependent on two parameters
$\alpha_{1}$ and $\alpha_{2}$ characterizing the positions of
observer and observable, respectively. Certainly, as projected to
QM, it becomes absolute since in this case only one parameter
$\alpha\approx 0;$ is measured.
\\
\\3) As indicated above, within the scope of QMFL it is obvious
that the states close to the origin singularity are always mixed,
being associated with the parameter $\alpha>0$ and hence with
nonzero entropy. At the same time, within QM one can have an
understanding (at least heuristically), in what way mixed states
are generated by the origin singularity.  Actually, in the
vicinity of the origin singularity, i.e. at Planck’s scale (where
the quantum-gravitational effects are considerable) the space-time
foam is formed \cite{Wheel},\cite{Foam} that from the
quantum-mechanical viewpoint is capable of generating only a mixed
state, the components of which are associated with metrics from
the space-time foam with certain probabilities arising from the
partition function for quantum gravitation
\cite{Wheel},\cite{Hawk}.

\section {Entropy Bounds and Entropy Density}
In the last few years Quantum Mechanics of black holes has been
studied under the assumption that GUR are valid
\cite{r6},\cite{r7},\cite{r15}. As a result of this approach, it
is indicated that the evaporation process of a black hole gives a
stable remnant with a mass on the order of the Planck’s $M_{p}$.
However, J.Bekenstein in \cite{bek1} has credited such an approach
as problematic, since then the objects with dimensions on the
order of the Planck length $\sim 10^{-33}cm$ should have very
great entropy thus making problems in regard to the entropy bounds
of the black hole remnants \cite{bek2}.
\\ In connection with this remark of J.Bekenstein \cite{bek1}
the following points should be emphasized:
\\ I. An approach proposed in \cite{sh1},\cite{sh3} and in the
present paper gives a deeper insight into the cause of high
entropy for Planck’s black hole remnants, namely: high entropy
density that by this approach at Planck scales takes place for
every fixed observer including that on a customary scale, i.e. on
$\alpha\approx 0$. In \cite{sh3} using the exponential ansatz
(\ref{U3}) it has been demonstrated how this density can increase
in the vicinity of the singularities with
\\
$$S_{in}=S_{0}^{0}\approx 0$$
\\
up to\\
$$S_{out}=S^{0}_{\frac{1}{4}}=-<ln[exp(-1/4)]\rho_{pure}>
=-<\ln(\rho^{*}(1/4))>
=\frac{1}{4}.$$
\\
when the initial state measured by the observer is pure.
\\ As demonstrated in \cite{sh1},\cite{sh3}, increase
in the entropy density  will be realized also for the observer
moving together with the information flow:
$S_{out}=S^{\frac{1}{4}}_{\frac{1}{4}}>S_{0}^{0}$, though to a
lesser extent than in the first case. Obviously, provided the
existing solutions for (\ref{U1}) are different from the
exponential ansatz (\ref{U3}), the entropy density for them
$S^{0}_{\alpha_{2}}$ will be increasing as compared to $S_{0}^{0}$
with a tendency of $\alpha_{2}$ to 1/4.
\\II. In works of J.Bekenstein, \cite{bek2} in particular,
a “universal entropy bound” has been used \cite{bek3}:
\begin{equation}\label{UBec}
S\leq 2\pi MR/\hbar,
\end{equation}
where $M$ is the total gravitational mass of the matter and $R$ is
the radius of the smallest sphere that barely fits around a
system. This bound is, however, valid for a weakly gravitating
matter system only. In case of black hole remnants under study it
is impossible to assume that on Planck scales we are concerned
with a weakly gravitating matter system, as in this case the
transition to the Planck’s energies is realized where
quantum-gravitational effects are appreciable, and within the
proposed paradigm  parameter $\alpha\approx 0$ is changed by the
parameter $\alpha>0$ or equally QM is changed by QMFL.
\\
\\III.This necessitates mentioning of the recent findings of R.Bousso
\cite{bou1},\cite{bou2}, who has derived the Bekenstein's
"universal entropy bound" for a weakly gravitating matter system,
and among other things in flat space, from the covariant entropy
bound \cite{bou3} associated with the holographic principle of
Hooft-Susskind \cite{hol1},\cite{hol2},\cite{hol3}.
\\ Also it should be noted that the approach proposed in
\cite{sh3},\cite{sh4} and in the present paper is consistent with
the holographic principle \cite{hol1}-\cite{hol3}. Specifically,
with the use of this approach one is enabled to obtain the entropy
bounds for nonblack hole objects of L.Susskind \cite{hol2}. Of
course, in (\cite{sh4}, section 6) and (\cite{sh3}, section 4) it
has been demonstrated, how a well-known semiclassical
Bekenstein-Hawking formula for black hole entropy may be obtained
using the proposed paradigm. Then we can resort to reasoning from
\cite{hol2}: “using gedanken experiment, take a neutral
non-rotating spherical object containing entropy $S$ which fits
entirely inside a spherical surface of the area $A$, and it to
collapse to black hole”. Whence \begin{equation}\label{USuss}
S\leq \frac{A}{4l^{2}_{p}}.
\end{equation}
Note also that the entropy density matrix
$S^{\alpha_{1}}_{\alpha_{2}}$ by its definition
\cite{sh1},\cite{sh3} falls into 2D objects, being associated with
$l^{2}_{min}\sim l^{2}_{p}$ \cite{sh4} and hence implicitly
pointing to the holographic principle.
\section {Conclusion}
Qualitative analysis performed in this work reveals that the
Information Loss Problem in black holes with the canonical problem
statement \cite{r16}-\cite{r18}   suggests in principle positive
solution within the scope of the proposed method – high-energy
density matrix deformation. Actually, this problem necessitates
further (now quantitative) analysis. Besides, it is interesting to
find direct relations between the described methods and the
holographic principle. Of particular importance seems a conjecture
following from \cite{bou2}:
\\is it possible to derive GUR for high energies
(of strong gravitational field) with the use of the covariant
entropy bound \cite{bou3} in much the same manner as R.Bousso
\cite{bou2} has developed the Heisenberg uncertainty principle for
the flat space?



\begin{thebibliography}{99}
%
%
\bibitem{sh1}
A.E.Shalyt-Margolin. Deformed density matrix, Density of entropy
and Information problem,[gr-qc/0307096].
%
%
\bibitem{sh2}
A.E.Shalyt-Margolin, Non-Unitary and Unitary Transitions in
Generalized Quantum  Mechanics and Information Problem Solving,
[hep-th/0309121]
%
%
\bibitem{sh3}
A.E.Shalyt-Margolin,Non-Unitary and Unitary Transitions in
Generalized Quantum  Mechanics, New Small Parameter and
Information Problem Solving, {\it Mod. Phys. Lett. A}, Vol 19, No.
5 (2004) pp.391-403,[hep-th/0311239].
%
%
\bibitem{sh4}
A.E.Shalyt-Margolin and J.G.Suarez,Quantum Mechanics at Planck's
scale and Density Matrix,{\it Intern.Journ.of Mod.Phys.}
D.12(2003)1265,[gr-qc/0306081]
%
%
\bibitem{sh5}
A.E.Shalyt-Margolin, Fundamental Length,Deformed Density Matrix
and New View on the Black Hole Information Paradox,[gr-qc/0207074]
%
%
\bibitem{sh6}
A.E.Shalyt-Margolin and A.Ya.Tregubovich, Generalized Uncertainty
Relations,Fundamental Length and Density Matrix,[gr-qc/0207068]
%
%
\bibitem{sh7}
A.E.Shalyt-Margolin and J.G.Suarez. Density Matrix and Dynamical
aspects of Quantum Mechanics with Fundamental Length,
[gr-qc/0211083]
%
%
\bibitem{sh8}
A.E.Shalyt-Margolin and J.G.Suarez,Quantum Mechanics of the Early
Universe and its Limiting Transition,[gr-qc/0302119]
%
%
\bibitem{r16}
S.Hawking,Breakdown of Predictability in Gravitational Collapse,
Phys.Rev.D14(1976)2460
%
%
\bibitem{r17}
S.Giddings,The Black Hole Information Paradox,[hep-th/9508151]
%
%
\bibitem{r18}
A.Strominger, Les Houches Lectures on Black Holes,
[hep-th/9501071].
%
%
\bibitem{pen1}
R.Penrose,Quantum Theory and Space - Time, in S.Hawking and
R.Penrose ,The Nature of Space and Time, Princeton University
Press,Princeton,New Jersey,1994
%
%
\bibitem{Helf}
A.D.Helfer,Do black holes radiate?, {\it Rept.Prog.Phys.}
66(2003)943,[gr-qc/0304042]
%
%
\bibitem{Hooft}
G.'t Hooft,Horizons,[gr-qc/0401027]
%
%
\bibitem{bek1}
J.D.Bekenstein,Black holes and information
theory,[quant-ph/0311049].
%
%
\bibitem{r1}
D. V. Ahluwalia,Quantum Measurement, Gravitation, and Locality,
Phys.Lett. B339 (1994)301
%
%
\bibitem{r2}
D.V.Ahluwalia,Wave-Particle duality at the Planck scale: Freezing
of neutrino oscillations, Phys.Lett. A275 (2000)31;Interface of
Gravitational and Quantum Realms, Mod.Phys.Lett. A17(2002)1135
%
%
\bibitem{r3}
M.Maggiore, The algebraic structure of the generalized uncertainty
principle,Phys.Lett.B319(1993)83
%
%
\bibitem{r4}
M.Maggiore,Quantum Groups,Gravity and Generalized Uncertainty
Principle, Phys.Rev.D49(1994)5182
%
%
\bibitem{r5}
S.Capozziello,G.Lambiase and G.Scarpetta, The Generalized
Uncertainty Principle from Quantum Geometry, Int.J.Theor.Phys. 39
(2000),15
%
%
\bibitem{r6}
R.Adler,P.Chen and D.Santiago,The Generalised Uncertainty
Principle and Black Hole Remnants,Gen.Rel.Grav.33(2001)2101; M.
Cavaglia, S. Das,How classical are TeV-scale black holes?
[hep-th/0404050]
%
%
\bibitem{Gar}
L.Garay,Quantum Gravity and Minimum Length
Int.J.Mod.Phys.A.A10(1995)145[gr-qc/9403008]
%
%
\bibitem{Heis}
W.Heisenberg,Uber den anschaulichen Inhalt der
quantentheoretischen Kinematik und Mechanik, Zeitsch.fur
Phys,43(1927)172
%
%
\bibitem{r7}
P.Chen and R.Adler, Black Hole Remnants and Dark
Matter,Nucl.Phys.Proc.Suppl.124(2003)103
%
%
\bibitem{Wheel}
J.A.Wheeler, Geometrodynamics and the Issue of the Final State,
in:{\it Relativity, Groups and Topology},p.317, eds.B.S. and C.M.
DeWitt.Gordon and Breach,N.Y.,1963
%
%
\bibitem{Foam}
L.Garay,Quantum Evolution in Spacetime Foam
Int.J.Mod.Phys.A14(1999)4079
%
%
\bibitem{Hawk}
S.Hawking,Spacetime Foam,Nucl.Phys.B114(1978)349
%
%
\bibitem{r15}
P.S.Custodio,J.E.Horvath,The Generalized Uncertainty
Principle,entropy bounds and black hole (non-)evaporation in
thermal bath,Class.Quant.Grav.20(2003)L197
%
%
\bibitem{bek2}
J.D.Bekenstein,Entropy bounds and black hole remnants,Phys.Rev.
D49 (1994) 1912, [gr-qc/9307035].

%
%
\bibitem{bek3}
J.D.Bekenstein,A universal upper bound on the entropy to energy
ratio for bounded systems,Phys.Rev. D23 (1981) 287.
%
%
\bibitem{bou1}
R.Bousso,Light-sheets and Bekenstein's bound,Phys.Rev.Lett. 90
(2003) 121302, [hep-th/0210295].
%
%
\bibitem{bou2}
R.Bousso,Flat space physics from holography,[hep-th/0402058].
%
%
\bibitem{bou3}
R.Bousso,A Covariant Entropy Conjecture,JHEP 9907 (1999) 004,
[hep-th/9905177];
%
%
\bibitem{hol1}
G.'t Hooft,Dimension reduction in quantum gravity,
[gr-qc/9310026]; The Holographic Principle (Opening
Lecture),[hep-th/0003004]
%
%
\bibitem{hol2}
L.Susskind,The world as a hologram,J.of Math. Phys.36 (1995) 6377
[hep-th/9409089];
%
%
\bibitem{hol3}
R.Bousso,The holographic principle,Rev.Mod.Phys. 74 (2002) 825
[hep-th/9905177]
%
%
\end{thebibliography}
\end{document}